\documentclass[a4paper]{article}
\usepackage{hyperref}
\usepackage{amsmath}
\usepackage{amsthm}
\usepackage{amsfonts}
\usepackage{mathrsfs}
\usepackage{graphicx}
\usepackage{epic,eepic}

\renewcommand\vec{\overrightarrow}

\def\noi{\noindent}

\newtheorem{theorem}{Theorem}
\newtheorem{lemma}{Lemma}
\newtheorem{proposition}{Proposition}
\newtheorem{definition}{Definition}
\newtheorem{corollary}{Corollary}
\theoremstyle{definition}
\theoremstyle{remark}

\title{Time-minimal control of dissipative two-level quantum systems: The integrable case}

\author{Bernard Bonnard\thanks{Institut de Math\'ematiques de Bourgogne, UMR CNRS 5584, 9 Avenue
        Alain Savary, BP 47 870 F-21078 DIJON Cedex FRANCE ({\tt bernard.bonnard@u-bourgogne.fr}).}
        \and Dominique Sugny\thanks{Institut Carnot de Bourgogne, UMR 5209 CNRS-Universit\'e de Bourgogne,
     9 Av. A. Savary, BP 47 870, F-21078 DIJON Cedex, FRANCE ({\tt
     dominique.sugny@u-bourgogne.fr}).}}

\begin{document}

\maketitle

\begin{abstract}
The objective of this article is to apply recent developments in
geometric optimal control to analyze the time minimum control
problem of dissipative two-level quantum systems whose dynamics is
governed by the Lindblad equation. We focus our analysis on the
case where the extremal Hamiltonian is integrable.
\end{abstract}

\noi\textbf{Keywords.} Optimal control, conjugate and cut loci,
quantum control\\

\noi\textbf{AMS classification.} 49K15, 70Q05\\

\pagestyle{myheadings}
\thispagestyle{plain}

\section{Introduction}
We consider a dissipative two-level quantum system whose dynamics
is governed by the Lindblad equation which takes the following
form in suitable coordinates $q=(x,y,z)$, i.e., in the coherence
vector formulation of density matrix \cite{altafini,schirmer}:
\begin{eqnarray}\label{eq1.1}
\dot{x}&=& -\Gamma x+u_2z \nonumber\\
\dot{y}&=& -\Gamma y-u_1z \\
\dot{z}&=& \gamma_--\gamma_+z+u_1y-u_2x .\nonumber
\end{eqnarray}
We refer to \cite{bonnardsugny} and \cite{sugnykontz} for the
details of the model. We recall that $x$ and $y$ are related to
off-diagonal terms of the density matrix of the system and $z$ to
the difference of population between the two states. In
(\ref{eq1.1}), $\Lambda=(\Gamma,\gamma_-,\gamma_+)$ is a set of
parameters such that $\Gamma\geq \gamma_+/2>0$ and $\gamma_+\geq
|\gamma_-|$ which describes the interaction of the two-level
system with the environment. More precisely, $\Gamma$ is the
dephasing rate and $\gamma_+$ and $\gamma_-$ are respectively
equal to $\gamma_{12}+\gamma_{21}$ and $\gamma_{12}-\gamma_{21}$
where the coefficients $\gamma_{12}$ and $\gamma_{21}$ are the
population relaxation rates. The control is the complex \emph{Rabi
frequency} $u=u_1+iu_2$ of the laser field which is assumed to be
in resonance with the frequency of the two-level system
\cite{boscainres}. The physical state belongs to the \emph{Bloch
ball}, $|q|\leq 1$, which is invariant for the dynamics
considered. If they are many articles devoted to optimal control
of quantum systems in the conservative case (see e.g.
\cite{boscaincharlot}), the dissipative case is still an open
problem.

The system can be written shortly as a \textit{bilinear system}
\begin{equation}\label{eq1.2}
\dot{q}=F_0(q)+u_1F_1(q)+u_2F_2(q)
\end{equation}
and in order to minimize the effect of dissipation, we consider
the time minimum control problem for which, up to a rescaling on
the set of parameters $\Lambda$, the control bound is $|u|\leq 1$.
The energy minimization problem with the cost $\int_0^T|u|^2dt$
where the time $T$ is fixed but the control bound is relaxed can
also be considered and shares similar properties.

A first step in the analysis of such systems is contained in
\cite{sugnykontz}. Assuming $u$ real, the problem can be reduced
to the time-optimal control of a two-dimensional system
\begin{eqnarray}\label{eq1.3}
\dot{y}&=& -\Gamma y-u_1z \\
\dot{z}&=& \gamma_--\gamma_+z+u_1y,\nonumber
\end{eqnarray}
with the constraint $|u_1|\leq 1$. For such a problem, the
geometric optimal control techniques for single-input
two-dimensional systems presented in \cite{boscainpiccoli} succeed
to make the time-optimal synthesis for every values of parameters
$(\Gamma,\gamma_-,\gamma_+)$.

In order to complete the analysis in the bi-input case, a
different methodology has to be applied and we shall make an
intensive use of techniques and results developed in a parallel
research project to minimize the transfer of a satellite between
two elliptic orbits (see \cite{bonnardsugny}). Such techniques are
two-fold.

First of all, the maximum principle will select extremal
trajectories, candidates as minimizers and solutions of an
Hamiltonian equation. A geometric analysis will identify the
symmetry group of the system and find suitable coordinates to
represent the Hamiltonian. A consequence of this analysis is the
fact that, in the case $\gamma_-=0$, the extremal system is
integrable and if $\Gamma=\gamma_+$, the problem can be in
addition reduced to a \emph{2D-almost Riemannian} problem on a
two-sphere of revolution for which a complete analysis comes from
\cite{bonnardtanaka}. We take advantage of this property to
analyze the general integrable case $\gamma_-=0$ using
continuation methods on the set of parameters, while the analysis
fits in the geometrical framework of \emph{Zermelo navigation
problems} \cite{bao}. This case corresponds to the physical
situation where the population relaxation rates are equal. The
mathematical tools presented in this article and a few numerical
simulations are sufficient to complete the analysis for
$\gamma_-=0$. The generic case where $\gamma_-\neq 0$ is treated
in a forthcoming article \cite{BMS} combining mathematical
analysis and intensive numerical computations.

Secondly, having selected extremal trajectories,
\emph{second-order conditions} using the variational equation and
implemented in the cotcot code \cite{BCT} allow to determine first
\emph{conjugate points} forming the conjugate locus which are
points where extremals cease to be locally optimal. Combined with
the geometric analysis, we can construct the \emph{cut locus}
which is formed by points where extremals cease to be globally
optimal.

The organization of this article is the following. In section
\ref{sec2}, we recall the maximum principle and the concept of
conjugate points associated to second-order optimality conditions.
In section \ref{sec3}, we present the geometric analysis of the
system, followed in section \ref{sec4} by a thorough analysis of
the so-called Grusin problem on a two-sphere of revolution. This
problem is generalized into a Zermelo navigation problem suitable
to our analysis. In section \ref{sec5}, we study the properties of
the extremals to analyze the optimal trajectories, combining
analytic and numerical methods.
\section{Geometric optimal control}\label{sec2}
\subsection{Maximum principle}
We consider the time minimum problem with fixed extremities $q_0$
and $q_1$. For a smooth system written $\dot{q}=F(q,u)$, with
$q\in\mathbb{R}^n$ and a control domain $U$ which is a compact
subset of $\mathbb{R}^m$, we have:
\begin{proposition}\label{prop:2.1}
If $(q,u)$ is an optimal control trajectory pair on $[0,T]$ then
there exists an absolutely continuous non-zero vector function
$(p,p_0)\in\mathbb{R}^n\times \mathbb{R}$ such that almost
everywhere on $[0,T]$, we have:
\begin{equation}\label{eq2.1}
\dot{q}=\frac{\partial H}{\partial p}(q,p,u),\
\dot{p}=-\frac{\partial H}{\partial q}(q,p,u)
\end{equation}
and
\begin{equation}\label{eq2.2}
H(q,p,u)=M(q,p)
\end{equation}
where $H(q,p,u)=\langle p,F(q,u)\rangle +p_0$, $p_0$ being a non
positive constant; $M$ is defined by $M(q,p)=Max_{u\in U}
H(q,p,u)$ and $M$ is zero everywhere.
\end{proposition}
\begin{definition}\label{def:2.1}
The mapping $H$ is called the \emph{pseudo-Hamiltonian}. A triple
$(q,p,u)$ solution of (\ref{eq2.1}) and (\ref{eq2.2}) is called
\emph{extremal} while the component $q$ is called an extremal
trajectory and $p$ is an adjoint vector.
\end{definition}
\textbf{Application:} We consider the time-minimum control problem
for a system of the form $\dot{q}=F_0(q)+\sum_{i=1}^mu_iF_i(q)$
with $m\geq 2$, $u=(u_1,\cdots,u_m)$, $|u|\leq 1$. We introduce
the Hamiltonian lifts $H_i=\langle p,F_i(q)\rangle$,
$i=0,1,\cdots,m$ of the vector fields $F_i$ and the set $\Sigma$
such that $H_i=0$ for $i=1,\cdots,m$. Then the maximization
condition (\ref{eq2.2}) leads to the following result.
\begin{proposition}\label{prop:2.2}
Outside $\Sigma$, an extremal control is given by
$u_i=H_i/\sqrt{\sum_{i=1}^m H_i^2}$, $i=1\cdots,m$ and extremal
pairs $z=(q,p)$ are solutions of the smooth true Hamiltonian
vector field $\vec{H}_r(z)$ with $H_r(z)=H_0(z)+(\sum_{i=1}^m
H_i^2(z))^{1/2}$.
\end{proposition}
\begin{definition}\label{def:2.2}
The surface $\Sigma$ is called the \emph{switching surface} and
the solutions of $\vec{H}_r(z)$ are called extremals of
\emph{order zero}. To be optimal, they have to satisfy $H_r(z)\geq
0$ and those with $H_r(z)=0$ are called \emph{abnormal}.
\end{definition}

An important but straightforward result is the following
proposition \cite{bonnardchyba}.
\begin{proposition}\label{prop:2.3}
Extremal trajectories of order zero correspond to singularities of
the \emph{end-point mapping}
\begin{equation}\label{eq2.3}
E^{q_0,T}:u\in L^\infty [0,T] \mapsto q(T,q_0,u)
\end{equation}
where $q(\cdot,q_0,u)$ denotes the response to $u$ with initial
condition $q_0$ and such that the control is restricted to the
$(m-1)$-sphere $|u|=1$.
\end{proposition}
\subsection{Second-order optimality conditions}
From proposition \ref{prop:2.3}, we can apply the concepts and
algorithms presented in \cite{BCT} to compute second-order
optimality conditions in the smooth case, when the control domain
$U$ is a manifold, $u$ being restricted to the unit sphere. The
framework
of this computation is next recalled.\\
\textbf{The concept of conjugate point:}

Since $U$ is a manifold, we may assume locally that
$U=\mathbb{R}^{m-1}$ and the maximization condition (\ref{eq2.2})
leads to
\begin{equation}\label{eq2.4}
\frac{\partial H}{\partial u}=0,\ \frac{\partial^2 H}{\partial
u^2}\leq 0 .
\end{equation}
Our first assumption is the \textit{strong Legendre-Clebsch condition}:\\
(H1) The Hessian $\partial^2H/\partial u^2$ is negative definite
along the reference extremal.\\
From the implicit function theorem, an extremal control can be
locally defined as a smooth function of $z=(q,p)$ and plugging $u$
into $H$ defines a smooth true Hamiltonian $H_r$.\\
Setting $M=\mathbb{R}^n$ and using Hamiltonian formalism, we
introduce:
\begin{definition}\label{def:2.3}
Let $z=(q,p)$ be a reference extremal defined on $[0,T]$. The
variational equation
\begin{equation}\label{eq2.5}
\delta\dot{z}=d\vec{H}_r(z(t))\delta z
\end{equation}
is called the \emph{Jacobi equation}. A \emph{Jacobi field} is a
non trivial solution $\delta z=(\delta q,\delta p)$. It is said to
be vertical at time $t$ if $\delta q(t)=0$.
\end{definition}

\noindent The following standard result is crucial.
\begin{proposition}\label{prop:2.4}
Let $L_0$ be the fiber $T_{q_0}^*M$ and let
$L_t=\exp[t\vec{H}_r(L_0)]$ be its image by the one parameter
subgroup generated by $\vec{H}_r$. Then $L_t$ is a Lagrangian
manifold whose tangent space at $z(t)$ is spanned by Jacobi fields
vertical at $t=0$. Moreover, the rank of the restriction to $L_t$
of the projection $\Pi : (q,p)\mapsto q$ is at most $(n-1)$.
\end{proposition}

\noindent We next formulate the relevant generic assumptions using
the end-point mapping.\\
\textbf{Assumptions}:
\begin{enumerate}
\item (H2) On each subinterval $[t_0,t_1]$, $0<t_0<t_1\leq T$, the
singularity of $E^{q(t_0),t_1-t_0}$ is of codimension one for
$u|_{[t_0,t_1]}$.
\item (H3) We are in the normal case $H_r\neq 0$.
\end{enumerate}

\noindent As a result, on each subinterval $[t_0,t_1]$ there
exists up to a positive scalar an unique adjoint vector $p$ such
that $(q,p,u)$ is extremal.
\begin{definition}\label{def:2.4}
We fix $q_0=q(0)$ and we define the \emph{exponential mapping}:
\begin{equation}\label{eq2.6}
\exp_{q_0}:(p(0),t)\mapsto
\Pi(\exp[t\vec{H}_r(q(0),p(0))])\nonumber
\end{equation}
where $p(0)$ is a $(n-1)$ dimensional vector, normalized with
$H_r=1$.
\end{definition}
\begin{definition}\label{def:2.5}
Let $z=(q,p)$ be the reference extremal on $[0,T]$. Under our
assumptions, a time $0<t_c\leq T$ is called \emph{conjugate} if
the mapping $\exp_{q_0}$ is not an immersion at $(p(0),t_c)$ and
the point $q(t_c)$ is said to be \emph{conjugate} to $q_0$. We
denote $t_{1c}$ the first conjugate time and $C(q_0)$ the
\emph{conjugate locus} formed by the set of first conjugate points
considering all extremal curves.
\end{definition}

\noindent We get:
\begin{theorem}\label{the:2.1}
Let $z(t)=(q(t),p(t))$ be a reference extremal on $[0,T]$
satisfying assumptions (H1), (H2) and (H3). Then the extremal is
optimal in the $L^\infty$-norm topology on the set of controls up
to the first conjugate time $t_{1c}$. Moreover, if $t\mapsto q(t)$
is one-to-one then it can be embedded into a set $W$, image by the
exponential mapping $\exp_{q(0)}$ of $N\times [0,T]$, where $N$ is
a conical neighborhood of $p(0)$. For $T<t_{1c}$, the reference
extremal trajectory is time minimal with respect to all
trajectories contained in $W$.
\end{theorem}

In order to get global optimality results, it is necessary to glue
together such \textit{micro-local sets}. We need to introduce the
following concepts.
\begin{definition}\label{def:2.6}
Given an extremal trajectory, the first point where it ceases to
be optimal is called the \emph{cut point} and taking all the
extremals starting from $q_0$, they will form the \emph{cut locus}
$Cut(q_0)$. The \emph{separating line} $SL(q_0)$ is formed by the
set of points where two minimizers initiating from $q_0$
intersect.
\end{definition}
\section{Geometric analysis of Lindblad equation}\label{sec3}
\subsection{Symmetry of revolution}
If we apply to system (\ref{eq1.1}) a change of coordinates
defined by a rotation of angle $\theta$ around the $z-$axis:
\begin{eqnarray}\label{eq3.1}
X &=& x\cos\theta+y\sin\theta \nonumber\\
Y &=& -x\sin\theta+y\cos\theta \nonumber \\
Z &=& z ,\nonumber
\end{eqnarray}
and a similar feedback transformation on the control:
\begin{equation}\label{eq3.2}
v_1=u_1\cos\theta+u_2\sin\theta ,~
v_2=-u_1\sin\theta+u_2\cos\theta ,\nonumber
\end{equation}
we obtain the system
\begin{eqnarray}\label{eq3.3}
\dot{X}&=& -\Gamma X+v_2Z \nonumber\\
\dot{Y}&=& -\Gamma Y-v_1Z \\
\dot{Z}&=& \gamma_--\gamma_+Z+v_1Y-v_2X .\nonumber
\end{eqnarray}
Hence, this defines a one dimensional symmetry group and by
construction $|u|=|v|$. Therefore, we deduce that the time-minimum
control problem (or the energy minimization problem) are invariant
for such an action. Using cylindric coordinates
\begin{equation}\label{eq3.4}
x=r\cos\theta,~ y=r\sin\theta, ~z=z \nonumber
\end{equation}
and the dual variables $p=(p_r,p_\theta,p_z)$, the Hamiltonian
$H_r$ takes the form
\begin{eqnarray}\label{eq3.5}
H_r&=& H_0+(H_1^2+H_2^2)^{1/2} \nonumber\\
&=& (-\Gamma
rp_r+(\gamma_--\gamma_+)p_z)+(z^2p_r^2+\frac{z^2}{r^2}p_\theta^2+
r^2p_z^2-4zrp_rp_z)^{1/2} .\nonumber
\end{eqnarray}
In particular, the Bloch ball is foliated by meridian planes
$\theta=\textrm{constant}$ in which the time-minimum synthesis is
the one associated to system (\ref{eq1.3}), where the control is
scalar and described in \cite{sugnykontz}. More precisely, we
have:
\begin{proposition}\label{prop:3.1}
For the time-minimum control, $\theta$ is a cyclic coordinate and
$p_\theta$ is a first integral of the motion. The sign of
$\dot{\theta}$ is given by $p_\theta$ and if $p_\theta=0$ then
$\theta$ is constant and the extremal synthesis for an initial
point on the z-axis is up to a rotation given by the synthesis in
the plane $\theta=0$. Up to a rotation, the control $u$ can also
be restricted to the single-input control $(u_1,0)$.
\end{proposition}
\begin{proof}
The proof is a generalization of the geometric situation
encountered in \cite{bonnardtanaka}. For the Hamiltonian vector
field $\vec{H}_r$, the points on the $z$-axis correspond to a
polar singularity and the extremals starting from the $z$-axis are
contained in meridian planes $\theta=\theta(0)$. Hence, $p_\theta$
is constant and extremal curves in the plane $\theta=\theta(0)$
are solutions of system (\ref{eq1.3}).
\end{proof}
\subsection{Spherical coordinates}
More properties can be seen using spherical coordinates:
\begin{equation}\label{eq3.6}
x=\rho\sin\phi\cos\theta,~ y=\rho\sin\phi\sin\theta,
~z=\rho\cos\phi \nonumber
\end{equation}
and a similar feedback transformation. We obtain the system:
\begin{eqnarray}\label{eq3.7}
\dot{\rho}&=& \gamma_-\cos\phi-\rho(\gamma_+\cos^2\phi+\Gamma\sin^2\phi) \nonumber\\
\dot{\phi}&=& -\frac{\gamma_-\sin\phi}{\rho}+\frac{\sin(2\phi)}{2}(\gamma_+-\Gamma)+v_2 \\
\dot{\theta}&=& -(\cot\phi)v_1 .\nonumber
\end{eqnarray}
and the corresponding Hamiltonian
\begin{eqnarray}\label{eq3.8}
H_r&=& [\gamma_-\cos\phi-\rho(\gamma_+\cos^2\phi+\Gamma\sin^2\phi)]p_\rho \nonumber\\
& & +
[-\frac{\gamma_-\sin\phi}{\rho}+\frac{\sin(2\phi)}{2}(\gamma_+-\Gamma)]p_\phi+\sqrt{p_\phi^2+
p_\theta^2\cot^2\phi} .\nonumber
\end{eqnarray}
In this representation, $(\phi,\theta)$ are the spherical
coordinates on the unit sphere of revolution around the z-axis:
$\theta$ is the angle of revolution and $\phi\in ]0,\pi[$ is the
angle of the meridian, $\phi=0,\pi$ correspond respectively to the
north and south poles.
\subsection{Lie brackets computations}
In order to complete the analysis, we immediately compute the Lie
brackets up to length 3 for the system written in Cartesian
coordinates as:
\begin{equation}\label{eq3.9}
\dot{q}=(G_0q+v_0)+u_1G_1q+u_2G_2q \nonumber
\end{equation}
where the $G_i$'s are the matrices
\begin{equation}\label{eq3.10}
G_0=\left(\begin{array}{ccc}-\Gamma&0&0\\0&-\Gamma&0\\0&0&-\gamma_+\end{array}\right),~
G_1=\left(\begin{array}{ccc}0&0&0\\0&0&-1\\0&1&0\end{array}\right),~
G_2=\left(\begin{array}{ccc}0&0&1\\0&0&0\\-1&0&0\end{array}\right)\nonumber
\end{equation}
and $^tv_0$ is the vector $(0,0,\gamma_-)$. It can be lifted into
a right-invariant control system on the semi-direct product
$GL(3,\mathbb{R})\times_S\mathbb{R}^3$ identified to the subgroup
of matrices of $GL(4,\mathbb{R})$ of the form:
\begin{equation}\label{eq3.11}
\left(\begin{array}{cc}1&0\\v&g\end{array}\right),~
v\in\mathbb{R}^3, g\in GL(3,\mathbb{R})\nonumber
\end{equation}
acting on the subset of vectors of $\mathbb{R}^4$:
$\left(\begin{array}{c}1\\q\end{array}\right),~q\in\mathbb{R}^3.$
To construct affine vector fields, we use the induced action of
the Lie algebra $(a,A)\cdot q=Aq+a$ and Lie brackets are given by
\begin{equation}\label{eq3.12}
[(a,A),(b,B)]=[Ab-Ba,AB-BA].\nonumber
\end{equation}
The control distribution is $D=Span\{G_1,G_2\}$ and we have:
\begin{equation}\label{eq3.13}
[G_1,G_2]=G_3=\left(\begin{array}{ccc}0&-1&0\\1&0&0\\0&0&0\end{array}\right).\nonumber
\end{equation}
In particular, we obtain that
$\{G_1,G_2\}_{A.L.}=\underline{so}(3)$ and hence the system on
$SO(3)$:
\begin{equation}\label{eq3.14}
\frac{dX}{dt}=(u_1G_1+u_2G_2)X\nonumber
\end{equation}
is controllable. For the linear action, it defines a controllable
system on the unit sphere. This action has however singularities:
\begin{itemize}
\item at 0, the orbit is 0.
\item the set on $\mathbb{R}^2$ where $G_1$ and $G_2$ are
collinear is the whole plane $z=0$ and restricted to the unit
sphere of revolution, it corresponds to the equator.
\end{itemize}
To analyze the effect of the drift term associated to dissipation,
we use
\begin{equation}\label{eq3.15}
[G_0,G_1]=(\Gamma-\gamma_+)\left(\begin{array}{ccc}0&0&0\\0&0&1\\0&1&0\end{array}\right),~
[G_0,G_2]=(\gamma_+
-\Gamma)\left(\begin{array}{ccc}0&0&1\\0&0&0\\1&0&0\end{array}\right)\nonumber
\end{equation}
and $[G_0,G_3]=0$. Moreover, we have
\begin{equation}\label{eq3.16}
[G_1,[G_0,G_1]]=2(\Gamma-\gamma_+)\left(\begin{array}{ccc}0&0&0\\0&1&0\\0&0&-1\end{array}\right),~
[G_2,[G_0,G_2]]=2(\gamma_+-\Gamma)\left(\begin{array}{ccc}1&0&0\\0&0&0\\0&0&-1\end{array}\right).\nonumber
\end{equation}
Those computations reveal the singularity at $\gamma_+=\Gamma$,
$\gamma_-=0$ that we describe in the next proposition.
\begin{proposition}\label{prop:3.2}
In the case $\gamma_-=0$, $\gamma_+=\Gamma$, the radial component
$\rho$ is not controllable and the time-minimum control problem is
an almost Riemannian problem on the two-sphere of revolution for
the metric in spherical coordinates $g=d\phi^2+\tan^2\phi
d\theta^2$ with Hamiltonian
$H=\frac{1}{2}(p_\phi^2+p_\theta^2\cot^2\phi)$.
\end{proposition}
\begin{proof}
If $\gamma_-=0$ and $\gamma_+=\Gamma$ then the first equation of
(\ref{eq3.7}) becomes $\dot{\rho}=-\Gamma \rho$ and $\rho$ is not
controllable. Hence, the time-minimal control problem reduces to
the problem of controlling $(\phi,\theta)$ in minimum time where
the associated true Hamiltonian is
$\sqrt{p_\phi^2+p_\theta^2\cot^2\phi}$. The time minimum problem
is equivalent to minimizing the length for the metric
$d\phi^2+\tan^2\phi d\theta^2$. According to Maupertuis principle,
we can replace the length by the energy with corresponding
Hamiltonian $\frac{1}{2}(p_\phi^2+p_\theta^2\cot^2\phi)$.
\end{proof}
\begin{definition}\label{def:3.1}
The almost Riemannian metric $g=d\phi^2+\tan^2\phi d\theta^2$ is
called the Grusin model on the two-sphere of revolution.
\end{definition}

\noindent Such a metric appears in quantum control in the
conservative case \cite{boscaincharlot} and a similar metric is
associated to orbital transfer \cite{bonnardcaillau}. It will be
analyzed in details in section \ref{sec4} since it is the starting
point of the analysis in the general case using a continuation
method on the set of parameters.

Another consequence of the previous computations is the
controllability properties of the system and the structure of
extremal trajectories.
\subsection{Controllability properties}
We recall that the Bloch ball $|q|\leq 1$ is invariant. Indeed,
introducing $\rho^2=|q|^2$, we get:
\begin{equation}\label{eq3.17}
\rho\dot{\rho}=-\Gamma(x^2+y^2)-\gamma_+z^2+\gamma_-z\leq 0
\end{equation}
which is strictly negative on the unit sphere except if
$x^2+y^2=0$, $|z|=1$ and $\gamma_+=|\gamma_-|$. Using the
representation (\ref{eq3.7}) of the system in spherical
coordinates, it is clear that we can control the angular variables
$\phi$ and $\theta$ if the controls are not uniformly bounded. If
$|u|\leq 1$ then we have restrictions depending upon the set of
parameters.

\noindent For $\gamma_-=0$, the system is homogeneous and $q=0$ is
a fixed point. The accessibility set in fixed time is with non
empty interior for a non-zero initial point, except in the case
$\gamma_+=\Gamma$ which corresponds to the Grusin model and for
which the time and energy minimization problem are equivalent.

\noindent The controllability properties for $|u|\leq 1$ are clear
in this case. Indeed, if $|\gamma_+-\Gamma|<2$ then we can
compensate the drift by feedback for the system on the two-sphere
of revolution, while it is not the case for $|\gamma_+-\Gamma|>2$.
\begin{proposition}\label{prop:3.3}
Let $q_0$ and $q_1$ be two points in the Bloch ball $|q|\leq 1$
such that $q_1$ is accessible to $q_0$. Then there exists a
time-minimum trajectory joining $q_0$ to $q_1$. Moreover, every
optimal trajectory is
\begin{enumerate}
\item either an extremal trajectory with $p_\theta=0$, contained
in a meridian plane, time-optimal solution of the two-dimensional
system (\ref{eq1.3}) where $u=(u_1,0)$.
\item either connection of smooth extremal arcs of order 0,
solutions of the Hamiltonian vector field $\vec{H}_r$ with
$p_\theta\neq 0$, while the only possible connections are located
in the equatorial plane $\phi=\pi/2$.
\end{enumerate}
\end{proposition}
\begin{proof}
The control domain is convex and the Bloch ball is compact. Hence,
we can apply the Filippov existence theorem \cite{lee}. In order
to get a regularity result about optimal trajectories, much more
work has to be done. This is due to the existence of a switching
surface $\Sigma$ : $H_1=H_2=0$ in which we can connect two
extremals arcs of order 0, provided we respect the
Erdmann-Weierstrass conditions at the junction, i.e., the adjoint
vector remains continuous and the Hamiltonian is constant. The set
$\Sigma$ can also contain singular arcs for which $H_1=H_2=0$
holds identically. Hence, we can have intricate behaviors for such
systems. In our case, the situation is simplified by the symmetry
of revolution.

Indeed, if $p_\theta=0$ then the singularities are related to the
classification of extremals in the single-input case, which is
described in \cite{sugnykontz}. We cannot connect an extremal with
$p_\theta\neq 0$ where $p_\theta$ is the global first integral
$xp_y-yp_x$ to an extremal where $p_\theta=0$ since the adjoint
vector has to be continuous.

Hence, the only remaining possibility is to connect two extremals
of order 0 with $p_\theta\neq 0$ at a point of $\Sigma$ leading to
the conditions $p_\phi=0$ and $p_\theta\cot\phi=0$ in spherical
coordinates. Since $p_\theta\neq 0$, one gets $\phi=\pi/2$. The
result is proved.
\end{proof}
\textbf{Remark}: The classification of extremal trajectories near
the equatorial plane is described in proposition \ref{prop:5.6}.
\section{The Grusin model on a two-sphere of revolution with
generalizations to Zermelo navigation problem}\label{sec4} The
Grusin model $g=d\phi^2+\tan^2\phi d\theta^2$ is a special case of
metrics of the form $d\phi^2+G(\phi)d\theta^2$ on a two-sphere of
revolution such that:
\begin{itemize}
\item (H1) $G'(\phi)\neq 0$ on $]0,\pi/2[$.
\item (H2) $G(\pi-\phi)=G(\phi)$ (reflective symmetry with respect
to the equator).
\end{itemize}
They appear in optimal control in the orbit transfer, smooth at
the equator or with a polar singularity, in quantum control and in
various geometric problems, e.g., Riemannian problems  on an
ellipsoid of revolution.  The importance of this control problem
has justified the recent analysis of \cite{bonnardtanaka} that we
complete next, using Hamiltonian formalism in order to make
generalizations. We first interpret the Grusin model as a
deformation of the round sphere.
\begin{definition}\label{def:4.1}
The standard homotopy between the Grusin model and the round
sphere on the two-sphere of revolution is
$g_\lambda=d\phi^2+G_\lambda(X)d\theta^2$ where
$G_\lambda(X)=\frac{X}{1-\lambda X}$, $X=\sin^2\phi$ and
$\lambda\in [0,1]$.
\end{definition}
By construction, the metric is analytic for $\lambda\in [0,1[$ and
for $\lambda=1$, we have the Grusin model with a pole of order 1
at the equator.\\
The first objective of this section is to show stability results
concerning such metrics. We have the following general result
\cite{bonnardtanaka}.
\begin{proposition}\label{prop:4.1}
Let $d\phi^2+G(\phi)d\theta^2$ be a smooth metric on a two-surface
of revolution. Then,
\begin{enumerate}
\item Extremals are solutions of the Hamiltonian
$H=\frac{1}{2}(p_\phi^2+\frac{p_\theta^2}{G(\phi)})$ and
arc-length parametrization amounts to restrict to $H=1/2$.
\item If $\psi$ is the angle of an unit-speed extremal with a
parallel then $p_\theta=\sqrt{G}\cos\psi$ is a constant and the
extremal flow is Liouville integrable with two commuting first
integrals $H$ and $p_\theta$.
\item The Gauss curvature is
$K=-\frac{1}{\sqrt{G}}\frac{\partial^2\sqrt{G}}{\partial \phi^2}$.
\end{enumerate}
\end{proposition}
\noindent We next make a complete analysis of the family of
metrics $g_\lambda$.
\subsection{Curvature analysis}
We have the following proposition.
\begin{proposition}\label{prop:4.2}
For the family of metrics $g_\lambda$, we have:
\begin{enumerate}
\item The Gauss curvature is
$K_\lambda=\frac{(1-\lambda)-2\lambda\
cos^2\phi}{(1-\lambda\sin^2\phi)^3}$.
\item $K'(\phi)=\frac{\lambda
\sin(2\phi)}{(1-\lambda\sin^2\phi)^4}[5(1-\lambda)-4\lambda\cos^2\phi]$.
\end{enumerate}
\end{proposition}
\noindent Hence $K(\phi)$ is non-constant and monotone non
decreasing from the north pole to the equator for $\lambda\in
]0,1/5]$, while for $\lambda\in ]1/5,1[$ it admits a minimum. For
$\lambda\in ]0,1[$, the curvature is maximum on the equator. The
limit case $\lambda=1$ corresponds to the Grusin case, for which
the curvature is negative everywhere and tends to $-\infty$ when
$\phi$ tends to $\pi/2$.
\subsection{Geometric properties}\label{sec4.2}
We next present the main properties of the extremal flow for a
metric on a two-sphere of revolution $g=d\phi^2+G(\phi)d\theta^2$
satisfying (H1) and (H2) where $g$ is smooth, except may be at the
equator where it can admit a pole of order one.

For such a family, we consider the smooth Hamiltonian
$H=\frac{1}{2}(p_\phi^2+\frac{p_\theta^2}{G(\phi)})$ and we
restrict extremal curves to the level set $H=1/2$. Fixing
$p_\theta$, the parameterized family of corresponding Hamiltonians
described the evolution of the $(p_\phi,\phi)$ variables as
solutions of a mechanical system for which
$V(\phi)=\frac{p_\theta^2}{G(\phi)}$ plays the role of potential.
For $p_\theta=0$, we get the meridian solutions. Hence, we can
assume $p_\theta\neq 0$. Using assumption (H1), the only
equilibrium point is for $\phi=\pi/2$ and $p_\phi=0$. This leads
to the equator solution in the regular case.

For the remaining trajectories, the level set $H=1/2$ is
sufficient to analyze the behaviors of $\phi$. Indeed, it is a
compact set, symmetric for the two reflections with respect to the
$\phi$-axis and the equator $\phi=\pi/2$ and defined respectively
by the two transformations: $p_\phi\mapsto-p_\phi$ and
$\phi\mapsto \pi-\phi$. Every trajectory is periodic and
$\psi=\pi/2-\phi$ oscillates periodically between $\psi_{max}$ and
$-\psi_{max}$. There is also a relation between the period of
oscillation $T$ and the amplitude $\psi_{max}$, depending upon
$p_\theta$.

By symmetry, every trajectory is defined by its restriction to a
quarter of period, that is the sub-arc starting from the equator
$\psi=0$ and reaching $\psi_{max}$. The trajectory starting from
$(\phi(0),p_\phi(0))$ and reaching $\pi-\phi(0)$ after passing
$\psi_{max}$ corresponds to a point rotating on the level set
$H=1/2$ and is chased by a point associated to the trajectory
starting from $(\phi(0),-p_\phi(0))$ and reaching $\pi-\phi(0)$.
They are distinct if $p_\phi(0)\neq 0$. Moreover, using the
assumption (H2), we deduce easily that for fixed $p_\theta$, the
extremals starting from $(\phi(0),\theta(0))$ with respectively
$p_\phi(0)$ and $-p_\phi(0)$ intersect with equal length on the
antipodal parallel. They are distinct if $p_\phi(0)\neq 0$. The
case $p_\phi(0)=0$ corresponds for an initial condition not on the
equator to tangential arrival and departure at parallels $\phi(0)$
and $\pi-\phi(0)$; $p_\phi(0)=0$ gives the equator solution in the
non singular case. For more details see \cite{bonnardcaillau2}.

The only difference in the singular case is that the equator is
not solution, and for trajectories departing from the equator the
extremals are always tangential to the meridian, while the first
return to the equator can be arbitrarily closed from the initial
point.

Finally, another obvious symmetry is a reflectional symmetry with
respect to the meridian obtained by changing $p_\theta$ into
$-p_\theta$.

As a consequence of this analysis, we deduce:
\begin{proposition}\label{prop:4.3}
Let $g=d\phi^2+G(\phi)d\theta^2$ be a metric on a two-sphere of
revolution, satisfying (H1) and (H2) and smooth except may be at
the equator where it can admit a pole of order one.
\begin{enumerate}
\item Then except the meridian and the equator solution in the
regular case, every extremal is such that $\psi=\pi/2-\phi$
oscillates periodically between two symmetric parallels. The first
return mapping to the equator is
\begin{equation}\label{eq4.1}
R:p_\theta\in ]0,\sqrt{G(\pi/2)}[\mapsto\Delta\theta(p_\theta),
\nonumber
\end{equation}
where $\Delta\theta$ is the corresponding $\theta-variation$ of
the extremal.
\item Assume $p_\theta\neq 0$ and $p_\phi(0) \neq 0$ then\\
a) Fixing $p_\theta$, changing $p_\phi(0)$ into $-p_\phi(0)$ gives
two distinct extremals with equal length intersecting on the
antipodal parallel.\\
b) Fixing $p_\phi(0)$ and changing $p_\theta$ into $-p_\theta$
gives two distinct extremals with equal length intersecting on the
opposite meridian.
\end{enumerate}
\end{proposition}
\subsection{Integrability}
For the family of metrics $g_\lambda$ which fit in the previous
geometric framework, we can be more precise and make a complete
analysis. For a fixed value of $\lambda$, the Hamiltonian is:
\begin{equation}\label{eq4.2}
H_\lambda=\frac{1}{2}(p_\phi^2+\frac{p_\theta^2}{G_\lambda(\phi)}),~
G_\lambda(\phi)=\frac{\sin^2\phi}{1-\lambda\sin^2\phi}\nonumber
\end{equation}
and corresponds for $\lambda=1$ to the Grusin case. Using
$\dot{\phi}=p_\phi$, we get:
\begin{equation}\label{eq4.3}
H_\lambda=\frac{1}{2}[\dot{\phi}^2+\frac{p_\theta^2(1-\lambda\sin^2\phi)}{\sin^2\phi}]
\nonumber
\end{equation}
which can be written
\begin{equation}\label{eq4.4}
H_\lambda=\frac{1}{2}[\dot{\phi}^2+p_\theta^2(\cot^2\phi+1-\lambda)].\nonumber
\end{equation}
Therefore, $H_\lambda=H_1+\frac{1}{2}p_\theta^2(1-\lambda)$ and
parameterized by arc-length: $H_\lambda=1/2$, one gets the level
set $H_1=\frac{1}{2}-\frac{1}{2}p_\theta^2(1-\lambda)$.

Hence the integration of the Grusin case gives the general
solution, and from the homotopy, the corresponding extremals fit
not only in the same geometric framework, but also have the same
transcendence.
\begin{lemma}\label{lemm:4.1}
The family of Hamiltonians $H_\lambda$ admits two first integrals
in involution for the Poisson bracket (independent of $\lambda$)
$p_\theta$ and $H_1=\frac{1}{2}[p_\phi^2+p_\theta^2\cot^2\phi]$.
\end{lemma}

We next outline the integration method in the Grusin case to
provide the computation of the first return mapping to the equator
$R$ obtained in \cite{bonnardtanaka}. We have
$H_1=\frac{1}{2}(\dot{\phi}^2+\nu\cot^2\phi),~\nu>0$ and fixing
the level set to $1/2$, we get:
\begin{equation}\label{eq4.5}
(\frac{d\phi}{dt})^2=\frac{1-(\nu+1)\cos^2\phi}{\sin^2\phi}.\nonumber
\end{equation}
Taking the positive branch, we must evaluate the following
expression
\begin{equation}\label{eq4.6}
\int\frac{\sin\phi d\phi}{(1-(\nu+1)\cos^2\phi)^{1/2}}=t.\nonumber
\end{equation}
To integrate, we use the relation
\begin{equation}\label{eq4.7}
\int\frac{\cos\phi
d\phi}{\sqrt{1-m^2\sin^2\phi}}=\frac{1}{m}\arcsin (m\sin\phi),
\nonumber
\end{equation}
to deduce the form of the component $\phi$ of the general
solution:
\begin{equation}\label{eq4.8}
\phi(t)=\arcsin[ \frac{1}{m}(\sin(mt+K))]+\pi/2. \nonumber
\end{equation}
To complete the integration, we write:
\begin{equation}\label{eq4.9}
\dot{\theta}=\frac{p_\theta}{\sin^2\phi}-\lambda p_\theta
\nonumber
\end{equation}
and we use the formula:
\begin{equation}\label{eq4.10}
\int\frac{dx}{1-a\sin^2x}=\frac{1}{\sqrt{1-a}}\arctan[\sqrt{1-a}\tan
x] \nonumber
\end{equation}
for $a<1$ with the relation $\cos\phi=-\frac{1}{m}\sin(mt+K)$. A
straightforward computation then leads to $\theta(t)$.
\begin{proposition}\label{prop:4.4}
For the family of metrics $g_\lambda$, we have:
\begin{equation}\label{eq4.11}
R(p_\theta)=\pi-\frac{\alpha \pi
p_\theta}{\sqrt{\alpha+1}\sqrt{\alpha+1+\alpha p_\theta^2}},~
\alpha=\frac{\lambda}{1-\lambda} .\nonumber
\end{equation}
In particular, if $\alpha>0$ then $R'(p_\theta)<0<R''(p_\theta)$
on $]0,\sqrt{G(\pi/2)}[$.
\end{proposition}

This property allows to evaluate conjugate and cut loci for the
family of metrics that we next describe \cite{bonnardtanaka}.
\subsection{Conjugate and cut loci}
We have:\\
\begin{theorem}\label{the:4.1}
\begin{itemize}
\item For $\lambda=0$ (round sphere), the conjugate and cut loci
of any point are reduced to the antipodal point.
\item For $0<\lambda<1$, the conjugate locus of a point different
from a pole is diffeomorphic to a standard astroid, while the cut
locus is a single branch of the antipodal parallel. Both are
symmetric with respect to the opposite meridian.
\item For $\lambda=1$ (Grusin case), the conjugate and cut loci of
a point different from a pole and not on the equator are as above.
For a point on the equator, the cut locus is the equator minus
this point and for the conjugate locus, the cusps on the equator
are transformed into folds at this point minus this point.
\end{itemize}
\end{theorem}
\textbf{Geometric interpretation}:\\
For the class of metrics $g_\lambda$, the situation is clear. For
the round sphere, all extremals starting from the equator
intersect at the same antipodal point and the first return mapping
is constant. For $0<\lambda\leq 1$, the first return mapping is
monotone, and in the singular case $R(p_\theta)\to 0$ as
$p_\theta\to +\infty$. Since the cut locus of a point of the
equator is formed by intersections with the equator of symmetric
extremals, in the homotopy, the cut locus is pinched into a point
for $\lambda=0$, while it is stretched into the whole equator in
the case $\lambda=1$.
\subsection{Zermelo navigation problem on the two-sphere of
revolution}\label{sec4.5} We introduce the following definition
for the Zermelo problem.
\begin{definition}\label{def:4.2}
A Zermelo navigation problem on the two-sphere of revolution is a
time-minimum problem of the form:
\begin{equation}\label{eq4.12}
\frac{dq}{dt}=F_0(q)+\sum_{i=1}^2u_iF_i(q),~ |u|\leq 1,\nonumber
\end{equation}
where the drift representing the current is of the form
$F_0^1(\phi)\frac{\partial}{\partial
\phi}+F_0^2(\phi)\frac{\partial}{\partial \theta}$ while $F_1$ and
$F_2$ form outside the equator an orthonormal frame for a metric
of the form $g=d\phi^2+G(\phi)d\theta^2$. It is called
reflectionaly symmetric with respect to the equator if
\begin{itemize}
\item (H1) $G'(\phi)\neq 0$ on $]0,\pi/2[$
\item (H2) $G(\pi-\phi)=G(\phi)$
\item (H3) $F_0^1(\pi-\phi)=-F_0^1(\phi)$,~ $F_0^2=0$.
\end{itemize}
It defines a Finsler geometric problem if $|F_0|<1$ for the metric
$g$.
\end{definition}

\noindent According to this classification, we have:
\begin{proposition}\label{prop:4.5}
Assume $\gamma_-=0$ and consider the system (\ref{eq3.7})
restricted to the two-sphere:
\begin{eqnarray}\label{eq4.13}
\dot{\phi}&=&\frac{\sin(2\phi)(\gamma_+-\Gamma)}{2}+v_2 \nonumber \\
\dot{\theta}&=& -(\cot\phi) v_1,~ |v|\leq 1. \nonumber
\end{eqnarray}
Then it defines a Zermelo navigation problem on the two-sphere of
revolution where the current is
$F_0^1=\frac{\sin(2\phi)}{2}(\gamma_+-\Gamma)$, the metric is
$g=d\phi^2+\tan^2\phi d\theta^2$ with a singularity at the equator
and the assumptions (H1), (H2) and (H3) are satisfied. The drift
can be compensated by a feedback when $|\gamma_+-\Gamma|<2$, which
defines a Finsler geometric problem on the sphere minus the
equator.
\end{proposition}
\textbf{Controllability analysis}\\
The amplitude of the current is $|\sin(2\phi)(\gamma_+-\Gamma)/2|$
and is maximum in the upper hemisphere for $\phi=\pi/4$, while it
is minimum at the north pole and at the equator. Hence, more
generally, we deduce the following proposition.
\begin{proposition}\label{prop:4.6}
For $|\gamma_+-\Gamma|>2$, the current can be compensated in the
north equator except in a band centered at $\phi=\pi/4$, hence
defining a Finsler geometric problem near the equator and near the
north pole.
\end{proposition}

The controllability analysis is straightforward and is related in
the north hemisphere to the scalar equation:
\begin{equation}\label{eq4.14}
\dot{\psi}=-\frac{\sin(2\psi)(\gamma_+-\Gamma)}{2}-v_2,~
\psi=\pi/2-\phi,~ |v_2|\leq 1.\nonumber
\end{equation}
Starting at $\psi=0$ with $v_2=-1$, to increase $\psi$ we meet a
barrier corresponding to the singularity of the vector field. For
instance, if $\gamma_+-\Gamma>0$ then we have a barrier when
$1=\sin(2\psi)(\gamma_+-\Gamma)/2,~ \psi\in ]0,\pi/2[$.
\section{The integrable case of two-level Lindblad equations}\label{sec5}
\subsection{The program}
We now proceed to the analysis of the general case of a two-level
Lindblad equation. The method is to start from the Grusin case and
then to consider perturbations. This program succeeds only if
$\gamma_-=0$, leading to extremal flows described by a family of
2D-integrable Hamiltonian vector fields.
\subsection{The integrable case}\label{sec5.2}
We observe that for $\gamma_-=0$, the true Hamiltonian simplifies
into:
\begin{equation}\label{eq5.1}
H_r=-\rho(\gamma_+\cos^2\phi+\Gamma\sin^2\phi)p_\rho+\frac{\sin(2\phi)(\gamma_+-\Gamma)}{2}p_\phi+
\sqrt{p_\phi^2+p_\theta^2\cot^2\phi}\nonumber
\end{equation}
and we immediately deduce:
\begin{proposition}\label{prop:5.1}
For $\gamma_-=0$, using the coordinate $r=\ln \rho$, the
Hamiltonian takes the form:
\begin{equation}\label{eq5.2}
H_r=-(\gamma_+\cos^2\phi+\Gamma\sin^2\phi)p_r+\frac{\sin(2\phi)(\gamma_+-\Gamma)}{2}p_\phi+
\sqrt{p_\phi^2+p_\theta^2\cot^2\phi}.\nonumber
\end{equation}
Hence $r$ and $\theta$ are cyclic coordinates and $p_r$,
$p_\theta$ are first integrals of the motion. The system is
Liouville integrable.
\end{proposition}

\noindent We have the following geometric interpretation.
\begin{proposition}\label{prop:5.2}
For $\gamma_-=0$, the Hamiltonian $H_r$ is associated if $p_r\leq
0$ to the problem of minimization of $r$, while the case $p_r\geq
0$ corresponds to the maximization of $r$, for $|u|=1$.
\end{proposition}
\begin{proof}
It is a consequence of the maximum principle. Another point of
view is to consider the end-point mapping $E:u\mapsto q(t,q_0,u)$.
If $u$ is restricted to the sphere $|u|=1$ then the solutions of
$H_r$ parameterize the singularities of the end-point mapping. The
case $p_r=0$ corresponds to singularities of the end-point mapping
for the system restricted to the two-sphere. In the extremum
problem of $r$, with fixed time, $p_r$ can be normalized to -1, 0
or 1, while the level sets are $H_r=h$. In the extremum problem of
time, the Hamiltonian is normalized to 0 or 1 for the minimum
case, and 0 or -1 for the maximum one. This is clearly equivalent
by homogeneity. Hence, this gives a dual point of view.
\end{proof}

\noindent In order to indicate the complexity of the problem, we
consider first the case of energy, which is equivalent to time
from Maupertuis principle, in the Grusin case.
\subsubsection{The case of energy}
In the normal case, the true Hamiltonian is
\begin{equation}\label{eq5.3}
H_r=-p_r(\gamma_+\cos^2\phi+\Gamma\sin^2\phi)+\frac{p_\phi\sin(2\phi)}{2}(\gamma_+-\Gamma)+
\frac{1}{2}(p_\theta^2\cot^2\phi+p_\phi^2).\nonumber
\end{equation}
We fix the level set to $h$ and using the relation
$p_\phi=\dot{\phi}+\frac{\sin(2\phi)}{2}(\Gamma-\gamma_+)$, one
gets $\frac{1}{2}\dot{\phi}^2+V(\phi)=h$ where $V(\phi)$ is the
potential:
\begin{equation}\label{eq5.5}
V(\phi)=-p_r(\gamma_+\cos^2\phi+\Gamma\sin^2\phi)-\frac{1}{2}\frac{\sin^2(2\phi)}{4}(\Gamma-\gamma_+)^2+
\frac{1}{2}p_\theta^2\cot^2\phi . \nonumber
\end{equation}
To integrate, we use:
\begin{equation}\label{eq5.7}
\frac{d\phi}{dt}=\pm \sqrt{2(h-V(\phi))}, \nonumber
\end{equation}
and we must evaluate an integral of the form:
\begin{equation}\label{eq5.8}
\int\frac{d\phi}{\sqrt{2(h-V(\phi))}}=\int
\frac{dX}{\sqrt{P(X)}}\nonumber
\end{equation}
where
\begin{equation}\label{eq5.9}
P(X)=4(1-X)[-X^3(\Gamma-\gamma_+)^2+X^2(\Gamma-\gamma_+)(2p_r+(\Gamma-\gamma_+))+
X(2h+2p_r\gamma_++p_\theta^2)-p_\theta^2] ,\nonumber
\end{equation}
with $X=\sin^2\phi$. This corresponds to an elliptic integral.
\subsubsection{The time-minimum case}
In the time-minimum case, the computations of the extremal curves
are more intricate because we cannot reduce the system to a
second-order differential equation. The geometric framework is
however neat because it is associated to a Zermelo navigation
problem.\\
\textbf{Computations}:\\
We set $Q=\sqrt{p_\phi^2+p_\theta^2\cot^2\phi}$ and the
Hamiltonian is restricted to a level set $\varepsilon$, where
$\varepsilon=0$ corresponds to the abnormal case and
$\varepsilon=+1$ to the normal case. This gives the following
relation:
\begin{equation}\label{eq5.10}
-(\gamma_+\cos^2\phi+\Gamma\sin^2\phi)p_r+(\gamma_--\Gamma)\frac{\sin(2\phi)}{2}p_\phi+Q=\varepsilon
\end{equation}
and fixing $p_r$ and $p_\theta$, the pair $\phi,p_\phi$ is
solution of the system:
\begin{eqnarray}\label{eq5.11}
\dot{\phi}&=& \frac{(\gamma_+-\Gamma)}{2}\sin(2\phi)+\frac{p_\phi}{Q} \nonumber \\
\dot{p_\phi}&=&
(\Gamma-\gamma_+)\sin(2\phi)p_r+(\gamma_+-\Gamma)\cos(2\phi)p_\phi+\frac{p_\theta^2\cos\phi}{Q\sin^3\phi}.
\end{eqnarray}
Hence $\dot{\phi}=0$ leads to:
\begin{equation}\label{eq5.12}
\frac{(\gamma_+-\Gamma)}{2}\sin(2\phi)Q+p_\phi=0.
\end{equation}
Using (\ref{eq5.10}), we deduce that $p_\phi$ is solution of a
polynomial equation of degree 2:
\begin{eqnarray}\label{eq5.13}
&
&p_\phi^2[(\gamma_+-\Gamma)^2\frac{\sin(2\phi)}{4}-1]-(\gamma_+-\Gamma)\sin(2\phi)
[\varepsilon+(\gamma_+\cos^2\phi+\Gamma\sin^2\phi)p_r]p_\phi
\nonumber \\& &+
[\varepsilon+(\gamma_+\cos^2\phi+\Gamma\sin^2\phi)p_r]^2-p_\theta^2\cot^2\phi=0
.
\end{eqnarray}
The discriminant of this polynomial is:
\begin{equation}\label{eq5.14}
\Delta=4[\varepsilon+(\gamma_+\cos^2\phi+\Gamma\sin^2\phi)p_r]^2+p_\theta^2\cot^2\phi[(\gamma_+-\Gamma)^2\sin^
2(2\phi)-4].
\end{equation}
From (\ref{eq5.12}), we deduce:
\begin{equation}\label{eq5.15}
p_\phi^2[(\gamma_+-\Gamma)^2\frac{\sin^2(2\phi)}{4}-1]=-(\gamma_+-\Gamma)^2\frac{\sin^2(2\phi)}{4}p_\theta^2\cot^2\phi.
\end{equation}
Hence the set $(\dot{\phi}=0)\cap (H_r=\varepsilon)$ is defined by
the relation:
\begin{equation}\label{eq5.16}
[\varepsilon+(\gamma_+\cos^2\phi+\Gamma\sin^2\phi)p_r]^2=p_\theta^2\cot^2\phi[1-\frac{(\gamma_+-\Gamma)^2}{4}\sin^2\phi].
\end{equation}

\noindent Therefore, we have:
\begin{lemma}\label{lemm:5.1}
\begin{enumerate}
\item If $(\gamma_+-\Gamma)^2\frac{\sin^2(2\phi)}{4}-1\neq 0$ and
$\Delta\geq 0$ then the level set $H_r=\varepsilon$ has two real
roots $p_\phi$ which are distinct if $\Delta>0$.
\item The intersection of $\dot{\phi}=0$ with the level set
$H_r=\varepsilon$ is given by $\Delta=0$ which can be written:
\begin{equation}\label{eq5.17}
[\varepsilon+(\gamma_+(1-X)+\Gamma
X)p_r]^2=\frac{p_\theta^2(1-X)}{X}[1-(\gamma_+-\Gamma)^2X(1-X)]
\end{equation}
where $X=\sin^2\phi$.
\end{enumerate}
\end{lemma}
\noindent \textbf{Extremals analysis}:

From the previous analysis, we deduce that there are two types of
extremal curves by considering the reduced system (\ref{eq5.11})
describing the evolution of
$(\phi,p_\phi)$.\\
\textbf{Compact case}:

It corresponds to the situation where the level sets
$H_r=\varepsilon$ define compact surfaces in the 2-plane
$(\phi,p_\phi)$. In this case, if the reduced system is without
singular point on the level set then the trajectory $t\mapsto
(\phi(t),p_\phi(t))$ is a periodic trajectory with period $T$.\\
The following lemma is clear.
\begin{lemma}\label{lemm:5.2}
The Hamiltonian $H_r$ is invariant for the transformation
$(\phi,p_\phi)\mapsto (\pi-\phi,-p_\phi)$.
\end{lemma}

\noindent A consequence of lemma \ref{lemm:5.2} is the following.
Assume that for $(p_\theta,p_r)$ fixed, a level set of $H_r$ is
such that it is compact without singular points and contains both
points $(\phi(0),p_\phi(0))$ and $(\pi-\phi(0),-p_\phi(0))$. In
this case, the trajectory starting from $(\phi(0),p_\phi^+(0))$
with $p_\phi^+(0)>0$ is periodic of period $T$ and has a second
crossing at the antipodal point $\pi-\phi(0)$ after a time $T/2$.
It is chased by a trajectory starting from $(\phi(0),p_\phi^-(0))$
where $p_\phi^+$ and $p_\phi^-$ are roots of (\ref{eq5.13}) for
$\phi=\phi(0)$ with a time delay of $T$ and reaches the antipodal
point $\pi-\phi(0)$ at time $T/2$.

Moreover, since the equations describing the evolutions of the
remaining variables $(r,\theta)$ are invariant for the central
symmetry: $(\phi,p_\phi)\mapsto (\pi-\phi,-p_\phi)$, we deduce
that after half a period $T/2$, we have, for the two extremals
starting respectively from $(\phi(0),p_\phi^+(0))$ and
$(\phi(0),p_\phi^-(0))$ while $(r(0),\theta(0)$ are identical, the
relations:
\begin{equation}\label{eq5.18}
r^+(T/2)=r^-(T/2),~ \theta^+(T/2)=\theta^-(T/2).
\end{equation}

\noindent This generalizes the analysis of section \ref{sec4.2}
for Riemannian metrics on the two-sphere, except that we replace
the reflectional symmetry by a central symmetry, leading to the
fact that half a period instead of a quarter of period is
necessary to construct the extremals.

We proved:
\begin{proposition}\label{prop:5.3}
If for fixed $(p_r,p_\theta)$, the level set $H_r=\varepsilon$ is
compact without singular point and has a central symmetry with
respect to $(\phi=\pi/2,p_\phi=0)$, then it contains a periodic
trajectory $(\phi,p_\phi)$ of period $T$ and if $p_\phi^\pm(0)$
are distinct then we have two distinct extremal curves $q^+(t)$,
$q^-(t)$ starting from the same point and intersecting with the
same length $T/2$ at a point such that $\phi(T/2)=\pi-\phi(0)$.
\end{proposition}
\textbf{Non-compact case}:

Non-compact level sets occur when $p_\phi\to\infty$. Using
(\ref{eq5.11}), the relation $\dot{\phi}=0$ gives the set $S$ of
solutions $\phi_S$ of
\begin{equation}\label{eq5.19}
(\gamma_+-\Gamma)^2\frac{\sin^2(2\phi)}{4}-1=0
\end{equation}
and we must have $|\Gamma-\gamma_+|\geq 2$. We deduce the
following proposition:
\begin{proposition}\label{prop:5.4}
If $|\Gamma-\gamma_+|\geq 2$ then we have extremal trajectories
such that $\phi$ is not periodic, i.e., $\dot{\phi}\to 0$,
$\phi\to \phi_S$ and $p_\phi\to \pm\infty$ when $t\to +\infty$
while $\dot{\theta}\to 0$ outside the equator.
\end{proposition}
\textbf{Geometric interpretation}:

The two types of extremals are related to the Zermelo navigation
problem of section \ref{sec4.5}.
\begin{itemize}
\item If $|\gamma_+-\Gamma|<2$ then the system restricted to the
two-sphere defines a Finsler geometric problem for which the
extremals of the Grusin case are deformed into extremals described
in proposition \ref{prop:5.3}.
\item If $|\gamma_+-\Gamma|\geq 2$ then we have two types of
extremals: periodic extremals occur in a band near the equator and
non periodic extremals occur when crossing the band around
$\phi=\pi/4$ where the current is maximal and asymptotic
properties of proposition \ref{prop:5.4} correspond to the barrier
phenomenon.
\end{itemize}

\noindent This will be clarify by the pictures of the final section.\\
\textbf{Small time optimal analysis at the equator}:\\
We make a singularity analysis of the extremals near a point of
the equator which can be identified to $\phi=\pi/2$ and
$\theta=0$. Following the techniques of \cite{bonnardchyba}, the
main point is to construct a normal form in order to compute the
small time optimal synthesis.
\begin{proposition}\label{prop:5.5}
Near the equator, the small time optimal synthesis is given by the
one near $(x,y,z)=(0,0,0)$ for the system:
\begin{eqnarray}\label{eq5.20}
\dot{x}&=& 1+\frac{(\gamma_+-\Gamma)}{\Gamma}y^2 \nonumber \\
\dot{y}&=& (\Gamma-\gamma_+)y+u_2 \\
\dot{z}&=& yu_1.\nonumber
\end{eqnarray}
\end{proposition}
\begin{proof}
The computation is straightforward. We set $\psi=\pi/2-\phi$ and
near $\psi=0$, we have
\begin{equation}\label{eq5.21}
\dot{r}\simeq -\Gamma+(\Gamma-\gamma_+)\psi^2,~ \dot{\psi}\simeq
(\Gamma-\gamma_+)\psi-u_2,~ \dot{\theta}\simeq -\psi u_1.\nonumber
\end{equation}
This gives the normal form by setting $x=-r/\Gamma$, $y=\psi$,
$z=\theta$ and using a feedback transformation preserving $|u|\leq
1$ and changing $(u_1,u_2)$ into $(-u_1,-u_2)$.
\end{proof}
This normal form preserves the integrability property and the
optimality analysis amounts to approximate the Hamiltonian $H_r$
near an equatorial point. Moreover, by homogeneity, it describes
the extremal behaviors not only near 0 but also along an
equatorial line identified to $(t,0,0)$.\\
From this normal form, we can deduce:
\begin{proposition}\label{prop:5.6}
For the system (\ref{eq5.20}), near 0, every small time optimal
trajectory is either
\begin{enumerate}
\item A trajectory in the meridian plane $z=0$ with $u_1=0$ such
that:\\
(i) If $\gamma_+-\Gamma<0$, an arc of the form BSB, i.e., a
concatenation of a bang arc $u_2=\pm 1$, a singular arc and a bang
arc,\\
(ii) If $\gamma_+-\Gamma>0$, a bang-bang arc BB,
\item or a smooth extremal of order 0 with $p_z\neq 0$.
\end{enumerate}
\end{proposition}
\begin{proof}
First of all, we analyze extremal curves contained in the plane
$z=0$ for which $u_1=p_z=0$. They correspond to the time-optimal
control problem in the 2D-plane $(x,y)$ with $|u_2|\leq 1$. They
are already analyzed in \cite{sugnykontz} using techniques of
\cite{boscainpiccoli}, but the computed normal form reveals also
the properties of the control. The line $S:t\mapsto (t,0)$ is a
singular trajectory and the corresponding control is $u_2=0$.
Moreover, this line is time minimizing if $\gamma_+-\Gamma<0$,
while it is time maximizing if $\gamma_+-\Gamma>0$. Other
extremals are defined by $u_2=\textrm{sign}(p_y)$. In case (i), a
bang arc can have a connection with the singular arc, with a
contact of order two with the switching surface. In contrast, the
connection is not possible in case (ii). To complete the analysis,
we observe that in case (i), every extremal curve which is of the
form BSB is optimal, while in the case (ii), every extremal curve
is bang-bang but for optimality, we have at most one switching.

In order to conclude the analysis, we must prove that an extremal
of order 0 with $p_z\neq 0$ cannot reach the singularity
$y=p_y=0$. The Hamiltonian $H_r$ takes the form
\begin{equation}\label{eq5.22}
H_r=p_x[1+\frac{(\gamma_+-\Gamma)}{\Gamma}y^2]+p_y(\Gamma-\gamma_+)y+\sqrt{p_y^2+p_z^2y^2}.\nonumber
\end{equation}
$H_r=\varepsilon$, with $\varepsilon=0,1$, gives the condition:
\begin{equation}\label{eq5.23}
p_x\frac{\gamma_+-\Gamma}{\Gamma}y^2+p_y(\Gamma-\gamma_+)y+\sqrt{p_y^2+p_z^2y^2}=0
.\nonumber
\end{equation}
This is clearly not possible, taking the Taylor expansions of $y$,
$p_y$:
\begin{equation}\label{eq5.24}
y(t)=at+o(t),~p_y(t)=bt+o(t)\nonumber
\end{equation}
of an extremal of order 0, with $p_z\neq 0$, reaching or departing
from the singularity. The result is proved.
\end{proof}

As a corollary, we have the following result to determine the
cut-locus in the case $|\gamma_+-\Gamma|<2$ where the time-minimal
control problem is related to a Finsler problem.
\begin{corollary}
Let $q^+(t)$ and $q^-(t)$ be two distinct extremals of order 0
with non-zero $p_\theta$ and intersecting with the same time $T$.
Then they cannot be optimal beyond the intersecting point.
\end{corollary}
\begin{proof}
The proof is standard. Assuming optimality beyond the intersecting
point, we can construct a broken minimizer which is an extremal of
order 0 with non-zero $p_\theta$. This contradicts proposition
\ref{prop:5.6}.
\end{proof}
\subsubsection{Numerical computations}
\begin{figure}[ht]
\centering
\includegraphics[width=0.6\textwidth]{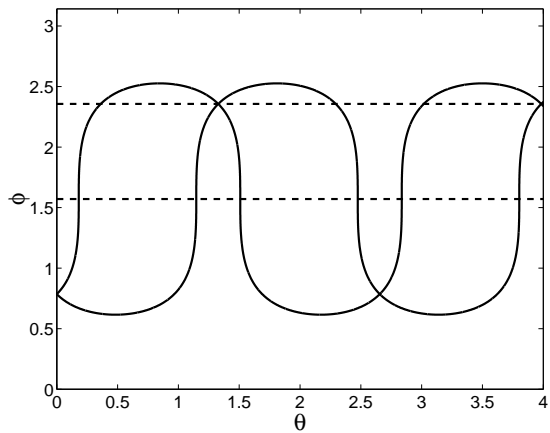}
\caption{Extremal trajectories for $\Gamma=2.5$ and $\gamma_+=2$.
Other parameters are taken to be $p_\phi(0)=-1$ and 2.33,
$\phi(0)=\pi/4$, $p_r=1$ and $p_\theta=2$. Dashed lines represent
the equator and the antipodal parallel located at $\phi=3\pi/4$.}
\label{fig1}
\end{figure}

\begin{figure}[ht]
\centering
\includegraphics[width=0.6\textwidth]{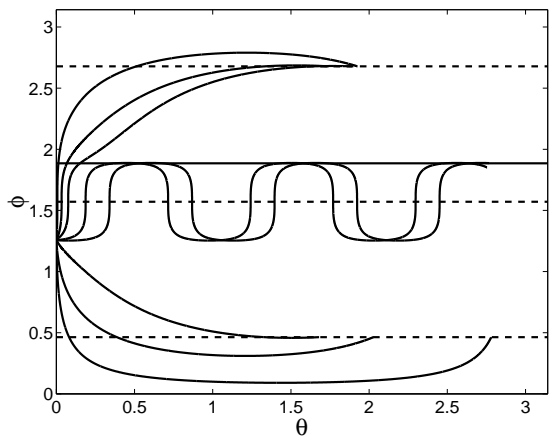}
\caption{Extremal trajectories for $\Gamma=4.5$ and $\gamma_+=2$.
Dashed lines represent the equator and the locus of the fixed
points of the dynamics given by equation (\ref{eq5.19}). The solid
line corresponds to the antipodal parallel. Numerical values of
the parameters are taken to be $\phi(0)=2\pi/5$, $p_\theta=8$ and
$p_r(0)=0.25$. The different initial values of $p_\phi$ are -50,
-10, 0, 2.637, 3, 5, 10 and 50.} \label{fig2}
\end{figure}

\begin{figure}[ht]
\centering
\includegraphics[width=0.6\textwidth]{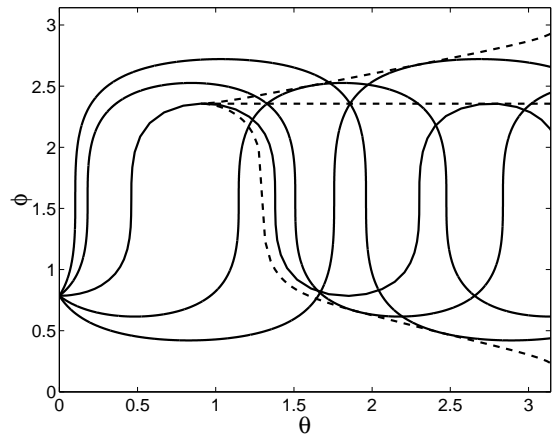}
\caption{Extremal trajectories for the Grusin model corresponding
to $\Gamma=\gamma_+=2$. The conjugate and cut loci are represented
in dashed lines. The cut locus in this case is the envelope of the
extremal curves. Numerical values are taken to be $p_\theta=2$ and
$p_r(0)=0.5$ and $\phi(0)=\pi/4$.} \label{fig3}
\end{figure}

\begin{figure}[ht]
\centering
\includegraphics[width=0.6\textwidth]{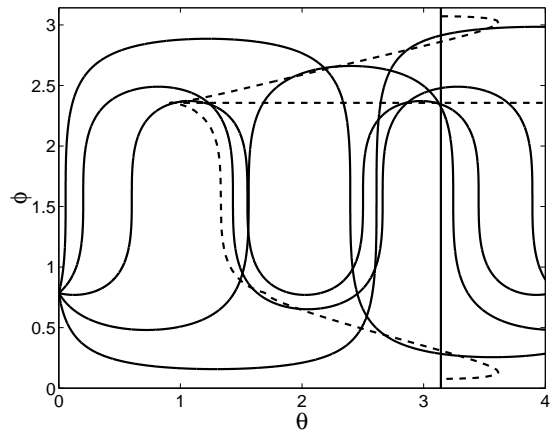}
\caption{Extremal trajectories for $\Gamma=2.5$ and $\gamma_+=2$.
The projection of conjugate locus is represented in dashed lines.
The horizontal dashed line is the line where two trajectories
intersect with the same length. Numerical values for the
parameters are taken to be $\phi(0)=\pi/4$, $p_r=0.5$ and
$p_\theta=2$.} \label{fig4}
\end{figure}

\begin{figure}[ht]
\centering
\includegraphics[width=0.6\textwidth]{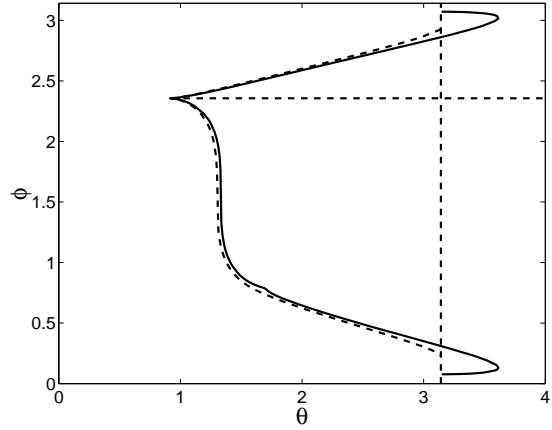}
\caption{Projection of conjugate locus in solid line for
$p_r=0.5$. The conjugate locus of the Grusin model corresponding
to $\gamma_+=\Gamma=2$ is represented in dashed lines. The
horizontal dashed line indicates the position of the cut locus for
the Grusin model. Dissipative parameters are taken to be
$\Gamma=2.5$ and $\gamma_+=2$, $p_\theta$ is equal to 2.}
\label{fig5}
\end{figure}

\begin{figure}[ht]
\centering
\includegraphics[width=0.6\textwidth]{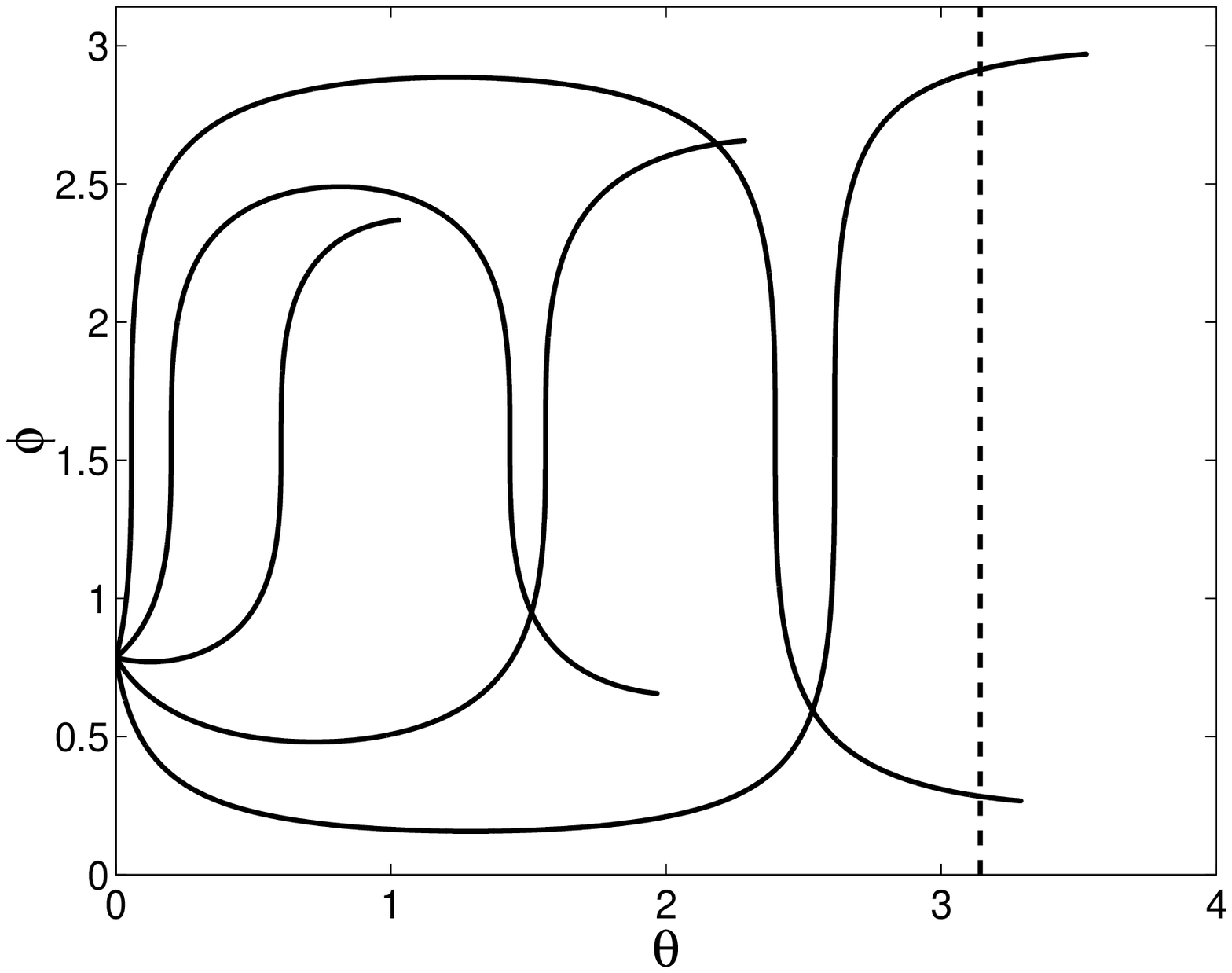}
\caption{Same as Fig. \ref{fig4} but the trajectories are plotted
up to the first conjugate point.} \label{fig6}
\end{figure}

We complete the analysis of section \ref{sec5} by a series of
numerical computations on extremal trajectories and on conjugate
loci (see section \ref{sec2}). All the computations are done in
the case of the time-minimum control problem but could be
equivalently done for the energy minimization problem. We use
numerical methods described in \cite{BCT}: an extremal is computed
and
we calculate along this trajectory the conjugate points.\\
\textbf{Extremal trajectories}:

We compute extremal trajectories for different values of the
dissipative parameters $\Gamma$ and $\gamma_+$. We represent in
figures \ref{fig1} and \ref{fig2} in the coordinates
$(\theta,\phi)$ the projection of the trajectories on the sphere
of radius 1. The variation of the radial coordinate $\rho$ can be
deduced straightforwardly and depends on the value of $\Gamma$ and
$\gamma_+$. Extremals can also be plotted in the plane
$(\phi,p_\phi)$ by using equation (\ref{eq5.10}). In figure
\ref{fig1}, two trajectories intersecting with the same cost on
the antipodal parallel are plotted. All other extremal
trajectories in the case $|\Gamma-\gamma_+|<2$ have the same
qualitative behavior. For the Grusin model on the sphere, we also
recall that the two trajectories which intersect on the cut locus
have opposite initial values of $p_\phi$, $\varepsilon$ and
$p_\theta$ being fixed. This is no more the case when $\Gamma\neq
\gamma_+$ since these initial values depend now on $\Gamma$ and
$\gamma_+$. The extremal trajectories for $|\Gamma-\gamma_+|>2$
are displayed in figure \ref{fig2}. We observe two types of
trajectories, the periodic and the aperiodic ones. The periodic
extremals have the same behavior as the ones of the case
$|\Gamma-\gamma_+|<2$. When $t\to +\infty$, the aperiodic
extremals have an asymptotic limit $\phi_S$ in $\phi$ solution of
the equation (\ref{eq5.19}). For given values of $p_\theta$ and
$\varepsilon$, this equation has two solutions symmetric with
respect to the equator. As can be seen in figure \ref{fig2}, we
have found two types of aperiodic extremals: trajectories monotone
in $\phi$ which do not cross the parallel $\phi=\phi_S$ and
trajectories passing by a maximum or a minimum in $\phi$ different
from $\phi_S$ and crossing this parallel. These different types of
behaviors can be determined by using equations (\ref{eq5.15})
and (\ref{eq5.16}).\\
\textbf{Conjugate locus}:

We next determine numerically the conjugate locus for different
set of parameters. We restrict the discussion to the case
$|\Gamma-\gamma_+|<2$. A similar study can be done for periodic
trajectories if $|\Gamma-\gamma_+|\geq 2$. For aperiodic
trajectories, it can be shown that they are locally optimal in the
sense that they do not have conjugate points for $t\in
[0,+\infty[$. Different conjugate loci are represented in figures
\ref{fig3}, \ref{fig4} and \ref{fig5} both for the Grusin model
and for a deformation of this model with the constraint
$|\Gamma-\gamma_+|<2$. The conjugate locus of the Grusin model on
the sphere is given in \cite{bonnardtanaka}. Here, in the case
$\Gamma=\gamma_+$, the drift vector $F_0$ being purely radial, the
projection of the conjugate locus is the same as the conjugate
locus of the Grusin model on the sphere. In particular, this
projection is independent of the value of $p_r$. Figure \ref{fig3}
displays both this conjugate locus and some extremal trajectories.
Starting from the Grusin model, we can then modify the difference
$\Gamma-\gamma_+$ and determine the corresponding deformation of
the conjugate locus. The evolution of the radial component being
more complicated (the radial coordinate is no more decoupled from
other coordinates if $\Gamma\neq \gamma_+$), the projection of the
conjugate locus depends on the value of $p_r$. A first comparison
between the two cases is given by figure \ref{fig4} where it can
be seen that the global structure of the extremals is nearly the
same. Figure \ref{fig5} displays the projection of a conjugate
locus on the plane $(\theta,\phi)$ corresponding to a particular
value of $p_r$. The conjugate locus of the Grusin model has been
added for the sake of comparison. We note that this locus is only
slightly modified when $\Gamma$ and $\gamma_+$ vary. The
trajectories of Fig. \ref{fig4} are plotted in Fig. \ref{fig6} up
to the first conjugate point.
\section*{Acknowledgments}
Agence Nationale de la recherche (ANR project CoMoc).

\end{document}